\documentclass[useAMS,usenatbib, usegraphicx]{mn2e}

\usepackage{amsmath}
\title[Eclipse mapping of RW Tri in the low luminosity state]{Eclipse mapping of RW Tri in the low luminosity state}
\author[A. V. Halevin, A. A. Henden]{A. V. Halevin$^{1}$\thanks{E-mail:
halevin@odessa-astronomy.org}, A. A. Henden$^{2}$\\ 
$^{1}$Department of Astronomy, Odessa National University, Odessa, 65014, Ukraine\\
$^{2}$American Association of Variable Star Observers, 49 Bay State Road, Cambridge, MA 02138, USA\\}

\begin{document}

\date{Accepted . Received ; in original form }

\pagerange{\pageref{firstpage}--\pageref{lastpage}} \pubyear{2008}

\maketitle

\label{firstpage}

\begin{abstract}
We analyzed the eclipse light curve of the nova-like star RW Tri in its low luminosity state. 
During approximately 150 days, RW Tri was about one magnitude fainter than in its usual state.
Our eclipse map shows that the brightness temperature in the disc ranges from 19000 K near the white dwarf 
to 8700 at the disc edge. 
For the inner parts of accretion disc, the radial temperature distribution is flatter than that predicted from the steady state models, and
for the outer parts, it is close to the $R^{-3/4}$ law. Fitting of the temperature 
distribution with one for the steady state disc model gives a mean accretion rate of (3.85$\pm$0.19)$\cdot$10$^{-9}$ M$_\odot$ year$^{-1}$. 
The hotspot in the disc is placed at a distance of 0.17$a$ from the white dwarf, where $a$ is the orbital separation.

\end{abstract}

\begin{keywords}
accretion, accretion discs - binaries: close - binaries: eclipsing - stars: individual:
RW Tri - novae, cataclysmic variables.
\end{keywords}

\section{INTRODUCTION}

RW Tri is a bright well known eclipsing nova-like system. It was discovered by Protitch in 1937 \citep{pro37}. \citet{wal63}
determined the orbital period to be 5.57~h. \citet{afr78} found that eclipse timings demand that the ephemeris has a cyclic term with
a period of 2777 or 4980 days. Different authors give different values of the system inclination angle $i$:
80$^\circ$ \citep{lon81}, 82$^\circ$ \citep{frk81}, 70.5$^\circ$ \citep{sma95}. \citet{frk81} found   
that the disc size is about 0.4$a$ where $a$ is the orbital separation.  

\citet{hos85} performed the first eclipse mapping of the system. They found that the
temperature of the inner part of accretion disc is about 40000 K. 
Also using the eclipse mapping technique, \citet{rpt92} determined the mass accretion rate to be 3$\cdot$10$^{-8}$ M$_\odot$ year$^{-1}$.

\citet{poo03} estimated the range for the primary and secondary stellar 
masses as 0.4  - 0.7 and 0.3 - 0.4  M$_\odot$ respectively. \citet{grp04}
using spectrophotometric data found that the mass accretion rate is about 10$^{-8}$ M$_\odot$
year$^{-1}$. Trigonometric parallax determination with the Hubble Space Telescope gave a distance of 341 pc to RW Tri \citep{mca99}.

\section{OBSERVATIONS}

In our paper we used AAVSO observations of RW Tri eclipse, obtained by Keith Graham with a Meade LX200 f/10 12" telescope and SBIG ST-9e CCD camera in {\it V} band. Observations were 
obtained during the low luminosity state (Fig.~1) on JD 2453672. The star brightness dropped from 12.6 to 13.7 mag in the {\it V} band for about 150 days. 
There are no outbursts observed during this state although the time interval between AAVSO visual measurements was sometimes longer than 25 days.
This shows that the accretion disc temperature did not drop below the instability limit.

The eclipse light curve is shown in Fig.~2. For our observations, the out-of-eclipse brightness of the system
was 13.74$\pm$0.06 mag and in the mid-eclipse the magnitude was 15.83$\pm$0.04 mag. One can see that the eclipse is deep and flickering on the light curve is
not significant. There is a small effect of the hotspot presence on the post eclipse light curve.  The light curve before the eclipse does not show the typical hump 
usually associated with a bright spot. In such a way we can conclude that the outer part of accretion disc is optically thin and the main source of
radiation in quiescent state is the central part of the accretion disc.

In this paper we used zero magnitude absolute fluxes, determined by \citet{bcp98} to prepare our observations for the eclipse mapping procedures.

\begin{figure}
%\imagei

\includegraphics[width=84mm]{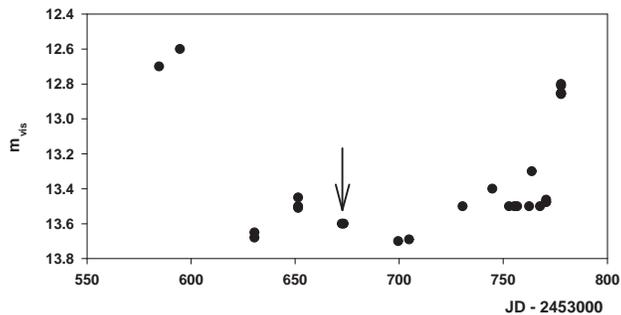}
\caption{Fragment of the AAVSO visual light curve of RW Tri. Eclipse observations are marked with an arrow.
\label{fig1}
} %% no full stop at the end
\end{figure}

%\begin{figure}
%\imagei
%\includegraphics[width=84mm]{fig2.eps}
%\caption{JD 2453672 V-band light curve of RW Tri, obtained with a Meade LX200 f/10 12" telescope and SBIG ST-9e CCD camera.} %% no full stop at the end
%\label{fig2}
%\end{figure}

\begin{figure}
%\imagei
\includegraphics[width=84mm]{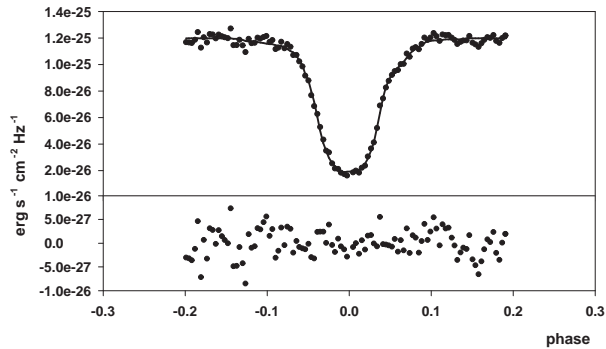}
\caption{Top: normalized light curve of RW Tri and the model fit. Fluxes were calculated using zero magnitude absolute fluxes, determined by \citet{bcp98}.
Below: residuals for the fit. One can see that large amplitude residuals
correspond to the flickering on out-of-eclipse parts of the light curve.} %% no full stop at the end
\end{figure}

\section{ECLIPSE MAPPING}

We applied a genetic algorithm eclipse mapping technique (see \citep{hal08} for detailed description) to calculate the eclipse map of the RW Tri accretion disc.
In our modification the accretion disc brightness is modeled with a distribution of radiating points in the orbital plane inside the Roche lobe of the primary star. 
Our technique looks for an optimal spatial distribution of the points to fit the observed eclipse light curve. The system flux is reconstructed here by summing of the
brightness of points visible at different phases. 

To remove smooth orbital brightness variations we used an polynomial approximation for the out-of-eclipse parts
of the light curve. After that we divided the eclipse light curve by the aproximation values and scaled the result with the polynomial value at
zero phase.

In our models we used system parameters taken from \citet{grp04} ($M_{wd}=0.7 M_\odot$, $M_{rd}=0.6 M_\odot$, $i$=75$^\circ$). 

One can see the normalized phase light curve of the eclipse with the fit and residuals
in Fig.~2. To build the map of accretion disc we used a model with 300 radiating points. The corresponding smoothed map for the brightness distribution 
in the accretion disc is in Fig. 3. The solid line inside the Roche lobe shows a ballistic stream trajectory. 

One can see that the brightest part of accretion disc is about 0.1$a$ in radius and the hotspot distance is about 0.17$a$. We consider this value as
the real size of accretion disc.  

\begin{figure}
\includegraphics[width=84mm]{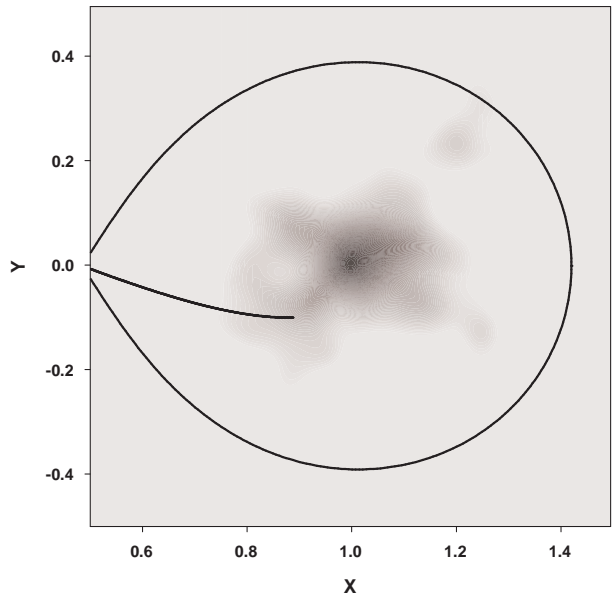}
\caption{Eclipse map for the JD 2453672 light curve of RW Tri calculated for the system parameters $q=0.86$ and $i=75^\circ$.
The solid line is the ballistic trajectory of the accretion stream. Spatial coordinates are in orbital separation units.} %% no full stop at the end
\end{figure}

\pagebreak

\begin{figure}
%\imagei
\includegraphics[width=84mm]{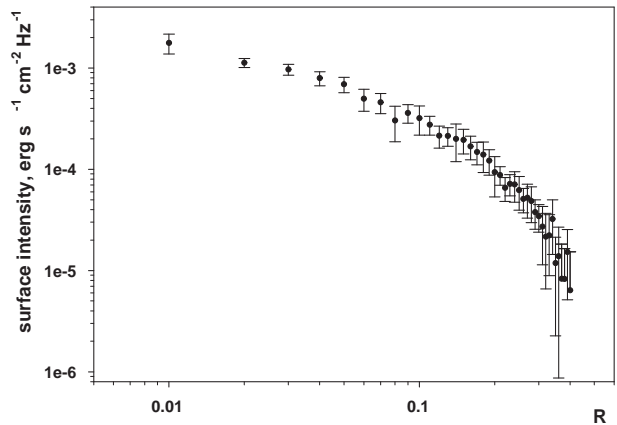}
\caption{Azimuthally averaged radial intensity distribution in accretion disc for Fig.3 eclipse map.} %% no full stop at the end
\end{figure}

\begin{figure}
%\imagei
\includegraphics[width=84mm]{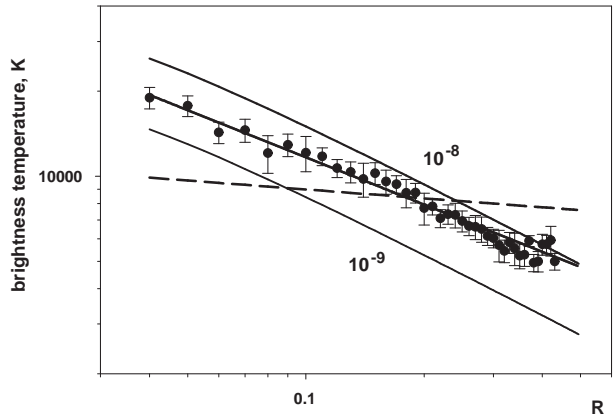}
\caption{Radial brightness temperature distribution in accretion disc. Solid lines are theoretical temperature distribution for
steady state disc in the case of 10$^{-8}$ and 10$^{-9}$ M$_\odot$ year$^{-1}$ mass accretion rate. 
Dashed line shows critical temperature above which gas is in steady accretion regime  \citep{war95}.} %% no full stop at the end
\end{figure}

\section{DISCUSSION}

Assuming the distance to the system to be 341 pc \citep{mca99}, we calculated the radial brightness temperature distribution in the disc and compared it with 
predictions of accretion disc models. Interstellar extinction $A_V$=7.8$\cdot$10$^{-4}$ mag $\cdot$ pc$^{-1}$ was taken using the value for the nearest
object from \citet{nks80}.
In Fig.5 the radial brightness temperature plot is shown. Here we compare the observed distribution with that predicted
for steady state solutions for accretion rates of 10$^{-8}$ and 10$^{-9}$ solar masses per year. Our fit of the temperature distribution with that
predicted from the steady state disc model gives the value $\dot M$ = (3.85$\pm$0.19)$\cdot$10$^{-9}$ M$_\odot$ year$^{-1}$. This result is close to 
that obtained from eclipse mapping by \citet{rpt92} value $\dot M$ = 3$\cdot$10$^{-9}$ M$_\odot$ year$^{-1}$.
We fitted the observed temperature distribution with the
function $T=T_0 R^{-b}$ and found $b = 0.56 (\pm 0.01)$. Our estimate of parameter $b$ is far 
from that predicted in the steady state model $3/4$ value. From Fig. 5 one can see that the temperature distribution consists of two different
parts: for $R$ less than 0.14$a$ and for $R$ greater than this radius.	
If we fit these parts separately, we obtain values $b = 0.52 \pm 0.04$ for $R<$0.14$a$ and $b = 0.74 \pm 0.03$ for $R>$0.14$a$.

Our mass accretion rate estimate is less than that determined by the \citet{hos85} value of
$\dot M$ = 10$^{-7.9}$ M$_\odot$ year$^{-1}$. According to their data the temperature in the disc does not drop below the critical value, above
which gas ramains in the steady state accretion regime, typical for classical nova-like systems. 

The dashed curve in Fig.5. shows
the critical temperature level \citep{war95}, calulated for the RW Tri system parameters. One can see that the temperature drops below critical value
immediately after the hotspot distance and, hence, the most probable accretion disc radius. It is enough for the accretion disc to remain in the steady state.

\section{CONCLUSIONS}

Using eclipse mapping techniques we calculated the radial brightness temperature distribution. For inner parts of accretion disc 
the slope of this distribution is close to the $R^{-1/2}$ law. For outer parts the temperature distribution corresponds to a steady state
$R^{-3/4}$ law. We estimated the mass accretion rate in the system as $\dot M$ = (3.85$\pm$0.19)$\cdot$10$^{-9}$ M$_\odot$ year$^{-1}$. 
Our results show that even during the low luminosity phase, disc remains in the hot steady state.

\section*{Acknowledgments}

We acknowledge with thanks the variable star observations from the AAVSO International Database contributed by observers worldwide and used in this research.

\bsp

\label{lastpage}

\end{document}